\documentclass[11pt,a4paper]{article}
\usepackage{jheppub}
\usepackage{verbatim}
\usepackage{lmodern}
\usepackage[T1]{fontenc}
\usepackage{mathtext,marvosym,textcomp}
\usepackage{indentfirst}
\usepackage{amsfonts}
\usepackage{mathtools}
\usepackage[usenames,dvipsnames,svgnames,table]{xcolor}
\usepackage{tikz}
\usetikzlibrary{arrows}
\usetikzlibrary{shapes.misc}

\tikzset{cross/.style={cross out, draw=black, minimum size=2*(#1-\pgflinewidth), inner sep=0pt, outer sep=0pt},
cross/.default={5pt}}
\usepackage[ normalem]{ulem}
\usepackage{tocloft}

\def\a{\alpha} \def\b{\beta} \def\g{\gamma} \def\d{\delta} \def\e{\epsilon} \def\h{\eta} \def\k{\kappa} \def\m{\mu} \def\n{\nu} \def\r{\rho} \def\s{\sigma}    \def\ff{\phi} \def\G{\Gamma} \def\W{\Omega}

\def\fr{\frac}  \def\dt{\partial}
\def\ph{\phantom}
\def\mc{\mathcal}
\def\mH{\mathcal{H}}
\def\mR{\mathcal{R}}
\def\tx{\tilde{x}}
\def\tdt{\tilde{\partial}}
\def\tm{\times}
\def\na{\nabla}

\def\cR{\hat{\cal R}}
\def\cG{\hat{\Gamma}}
\def\cN {\hat{\nabla}}
\def\XX{\mathbb{X}}

\title{Classical Yang-Baxter equation from \boldmath $\beta$-supergravity}
\author[a,c]{Ilya Bakhmatov}
\author[b,c]{and Edvard T. Musaev}
\affiliation[a]{Asia Pacific Center for Theoretical Physics,\\
	Postech, Pohang 37673, Korea}
\affiliation[b]{Moscow Institute of Physics and Technology,\\
	Institutskii per. 9, Dolgoprudny, 141700,  Russia}
\affiliation[c]{Kazan Federal University, Institute of Physics,\\
	Kremlevskaya 16a, Kazan, 420111, Russia}
\emailAdd{ilya.bakhmatov@apctp.org}
\emailAdd{musaev.et@phystech.edu}
\abstract{Yang-Baxter deformations of superstring $\s$-models have recently inspired a supergravity solution generating technique. Using the open/closed string map and a Killing bi-vector as a deformation parameter, new solutions can be built, such that the (generalised) supergravity field equations were conjectured to always reduce to the classical Yang-Baxter equation (CYBE)~\cite{Bakhmatov:2017joy}. In this work we provide a proof of this conjecture, using a systematic approach based on the so-called $\b$-supergravity, which is a dynamical theory for the field $\b^{mn}$ instead of the NSNS 2-form $b_{mn}$.}
\keywords{Supergravity Models, Space-Time Symmetries, String Duality}
\arxivnumber{1811.09056}

\begin{document}

\maketitle

\section{Introduction}

Recently it has been proposed that for any solution of $\mathcal{N} = 2$, $d = 10$ supergravity with isometries, there exists a very simple deformation, such that the Classical Yang-Baxter equation (CYBE) for the deformation parameter implies that the deformed background is a solution to the supergravity field equations~\cite{Bakhmatov:2017joy, Bakhmatov:2018apn}. For a supergravity solution specified by the metric $G_{mn}$, dilaton $\Phi$ and possibly some Ramond-Ramond (RR) fields, but with vanishing Kalb-Ramond tensor $B_{mn} = 0$, the deformation produces a solution given by $g_{mn}$, $b_{mn}$, $\phi$, according to the transformation rules:
	\begin{equation}
	\label{op-cl-map}
	g_{mn} + b_{mn} = ( G^{-1} + \b )^{-1}{}_{mn}, \qquad e^{-2\phi} |\det g_{mn}|^{1/2} = e^{-2\Phi} |\det G_{mn}|^{1/2}.
	\end{equation}
Here the deformation parameter $\b^{mn}$ is an antisymmetric tensor, which is constructed from the Killing vectors $k_a^m\dt_m$ of the original background metric $G_{mn}$:
	\begin{equation}
	\label{ansatz}
	\b^{mn} = r^{ab} k_a^m k_b^n, \qquad r^{ab} = -r^{ba}.
	\end{equation}
The conjecture put forward in~\cite{Bakhmatov:2017joy} was that the supergravity equations of motion for the deformed fields $g_{mn}$, $b_{mn}$, and $\phi$ are satisfied, as soon as the matrix $r^{ab}$ (the $r$-matrix) obeys the CYBE:
	\begin{equation}\label{CYBE}
	f_{de}{}^{[a} r^{b|d|}r^{c]e}=0,
	\end{equation}
where $f_{ab}{}^c$ are structure constants of the isometry algebra, $[k_a,k_b]=f_{ab}{}^c k_c$. 
	
Emergence of the CYBE from a purely gravitational theory is remarkable, given the prominent role played by the CYBE in integrable systems~\cite{Jimbo:1989qm}. The conjecture outlined above can be viewed as a concrete realisation of the earlier proposal of a gravity/CYBE correspondence~\cite{Matsumoto:2014cja},~\cite{Matsumoto:2014nra}. It was suggested in~\cite{Bakhmatov:2017joy} that the deformation~\eqref{op-cl-map} can be used as a solution generating technique for arbitrary supergravity solutions with isometries. Although in this article we will concentrate exclusively on the NSNS sector of type~II supergravity, note that the complete recipe for the deformation including the RR fields can be found in~\cite{Bakhmatov:2017joy}.
	
The map~\eqref{op-cl-map} was originally introduced by Seiberg and Witten~\cite{Seiberg:1999vs} in their seminal study of open strings on D-branes in the background fields $g_{mn}, b_{mn}$, which belong to the closed string spectrum. It was demonstrated that the open string can effectively be described as propagating on a non-commutative spacetime with the metric $G_{mn}$ and non-commutativity parameter $\beta^{mn}$ (proportional to the quantum commutator $[x^m,x^n]$). We will thus refer to the field redefinition~\eqref{op-cl-map} as the open/closed string map.

Introduction of the solution generating technique of~\cite{Bakhmatov:2017joy} has built upon the earlier observation~\cite{Araujo:2017jkb, Araujo:2017jap} that the open/closed string map has the same effect on supergravity solutions as the Yang-Baxter deformation, developed in~\cite{Klimcik:2002zj, Klimcik:2008eq, Delduc:2013qra, Kawaguchi:2014qwa} for classically integrable superstring $\sigma$-models in the coset formalism. From the supergravity viewpoint, however, the solution generation prescription of~\cite{Bakhmatov:2017joy} is more generic in that it is applicable to coset and non-coset geometries alike. The fundamental difference is that in the $\sigma$-model approach the $r$-matrix solution to the CYBE has to be put in by hand, and this step is necessary for the deformed solution to preserve the classical integrability of the $\s$-model. In the approach of~\cite{Bakhmatov:2017joy} this logic is essentially reversed: the deformation parameter is given by~\eqref{ansatz} with an arbitrary $r$-matrix, and the CYBE emerges after imposing the supergravity field equations on the deformed solution. Presence or absence of superstring integrability in the initial background plays no role for the workings of the deformation. Whether emergence of the CYBE after the deformation hints to integrability of any kind has yet to be seen.
	
Various classes of Yang-Baxter deformations have been linked to T-duality-shift-T-duality (TsT) transformations~\cite{Osten:2016dvf}, or more generally non-abelian T-dualities~\cite{Hoare:2016wsk, Borsato:2016pas, Borsato:2017qsx, Lust:2018jsx} and $O(d,d)$ transformations~\cite{Sakamoto:2017cpu,Fernandez-Melgarejo:2017oyu,Sakamoto:2018krs,Araujo:2018rho}. 
In certain cases, the Yang-Baxter deformation may lead to a background that is not, strictly speaking, a solution to supergravity. Such backgrounds have been interpreted as resulting from T-duality in non-isometric direction~\cite{Hoare:2015wia, Orlando:2016qqu}. Although not supergravity solutions, they were shown to satisfy the field equations of generalised supergravity~\cite{Arutyunov:2015mqj, Wulff:2016tju}, where the generalisation consists in certain additional terms that depend on a Killing vector field $I^m$. The latter is related to the deformation (non-commutativity) parameter~\cite{Araujo:2017jap, Bakhmatov:2017joy}:
	\begin{equation}\label{I}
	I^m = \nabla_k \b^{km}.
	\end{equation}
Whenever this happens to be zero, the generalised field equations reduce to those of the usual type II supergravity; otherwise one has to deal with the generalised equations and the CYBE emerges from there as well. 

Most recently, a generalisation of the open/closed string map~\eqref{op-cl-map} resulting from non-abelian T-dualities was proposed in the sigma-model setup~\cite{Borsato:2018idb}. This paper has extended the previous works, which rely on the coset formulation of the sigma-model action, to generic sigma-models with isometries. Assuming the CYBE for the deformation parameter it was proven that kappa-symmetry is preserved under the deformation, which implies that the background fields satisfy the (generalised) supergravity field equations~\cite{Wulff:2016tju}.

In the supergravity framework of~\cite{Bakhmatov:2017joy}, the conjecture that the CYBE appears from the (generalised) supergravity field equations after the deformation~\eqref{op-cl-map} was supported by explicit examples of coset and non-coset geometries alike, such as $AdS_2 \times S^2$ and Schwarzschild spacetimes. More examples of generated solutions both to standard and to generalised supergravity have appeared in~\cite{Bakhmatov:2017joy, Bakhmatov:2018apn, Araujo:2018rbc}. In the present article we will give a proof of the conjecture at the level of supergravity action and field equations. This will enable us to highlight the relevant dual field theory for this problem, the $\b$-supergravity. The use of supergravity language will help us pose certain unsolved problems that cannot be addressed in the string sigma-model formalism: the generalised version of $\b$-supergravity, and $d=11$ generalization of the whole Yang-Baxter deformation narrative.

Proof of the conjecture at the supergravity level can be achieved, in principle, by making the field redefinition~\eqref{op-cl-map} directly in the supergravity field equations for the transformed fields $g_{mn}, b_{mn}, \phi$. One can then expand the equations in powers of $\beta^{mn}$, use the Killing bi-vector ansatz~\eqref{ansatz}, and see if the structure of the CYBE emerges. Such perturbative approach to proving the conjecture was undertaken recently in~\cite{Bakhmatov:2018apn}, where it was shown that up to the third order in powers of $\beta$ the supergravity field equations reduce to the CYBE. The complete non-perturbative proof for an arbitrary initial metric $G_{mn}$  was still lacking.
	
It turns out, however, that the field redefinition~\eqref{op-cl-map} can be performed consistently in the full type II supergravity action. This has been done explicitly in~\cite{Andriot:2013xca} and the resulting theory whose dynamical fields are the metric $G_{mn}$, the bi-vector field $\b^{mn}$, and the dilaton $\Phi$ is called $\b$-supergravity. Initially the interest in rewriting of the theory in such a way stemmed from the goal of finding a supergravity description for non-geometric backgrounds characterised by the $Q$-flux~\cite{Andriot:2011uh,Andriot:2012wx}. In the framework of $\b$-supergravity the $Q$-flux is simply given by $Q_m{}^{pq}=\nabla_m\b^{pq}$, and in general can be related to the torsion of the Weitzenb\"ock connection in Double Field Theory~\cite{Geissbuhler:2013uka,Andriot:2014uda}.

Double Field Theory (DFT) provides the most transparent understanding of the structure of $\b$-supergravity.  DFT is a field theory on the doubled spacetime that incorporates the usual supergravity and is explicitly covariant under the $O(d,d)$ symmetry group coming from the string theory T-duality~\cite{Hohm:2010jy,Hohm:2010pp} (see~\cite{Berman:2013eva,Hohm:2013bwa} for review; earlier applications of DFT to Yang-Baxter deformations include~\cite{Sakamoto:2017wor,Sakamoto:2017cpu,Fernandez-Melgarejo:2017oyu,Sakamoto:2018krs}). Dynamical fields of the theory are the T-duality invariant dilaton $d$ and the so-called generalised metric   
\begin{equation}
\mH \in \fr{O(10,10)}{O(1,9)\times O(1,9)}.
\end{equation}
Such coset element can be parametrised by the metric $g_{mn}$ and the Kalb-Ramond field $b_{mn}$, which gives the usual field content of the NS-NS sector of supergravity. Alternatively, the same element can be parametrised by the fields $G_{mn}$ and $\b^{mn}$, in which case DFT reduces to $\beta$-supergravity. At the level of fields this reparametrization of the coset element is precisely the open-closed string map \eqref{op-cl-map}.

This allows to make a formal transition from the standard description of type~II supergravity with dynamical fields $g_{mn}, b_{mn}, \phi$, which in this context is referred to as the $b$-frame of DFT, to the description in terms of the fields $G_{mn}, \beta^{mn}, \Phi$ in the $\beta$-frame. The dynamical equations in the $\beta$-frame can be used to derive the CYBE under the assumption of the ansatz~\eqref{ansatz}, as we show below.
	
This paper is organised as follows. In section \ref{sec2} a short review of $\b$-supergravity is given, followed by the proof of the conjecture. In section \ref{sec3} we discuss possible applications and consequences of the technique.

\section{The gravity/CYBE correspondence}
\label{sec2}

Starting with a solution to the ordinary supergravity $G_{mn}, \Phi$ with vanishing $B_{mn}$, one deforms it by introducing a 2-form $\beta^{mn}$. We interpret the resulting configuration $G_{mn}, \beta^{mn}, \Phi$ as a supergravity background in the $\beta$-frame. The DFT construction then implies that the background $g_{mn}, b_{mn}, \phi$ resulting from the open/closed string map is a solution to the standard $b$-frame supergravity field equations. The crucial point of this procedure is that the same fields $G_{mn}, \Phi$ must furnish a solution to the usual supergravity before the deformation, as well as to $\beta$-supergravity, when accompanied by a deformation field $\beta^{mn}$. This restricts possible deformations by effectively imposing the CYBE. Solving it for a given background $G_{mn}, \Phi$ will determine all possible deformations of this background. This can be summarised with a diagram of Figure 1:

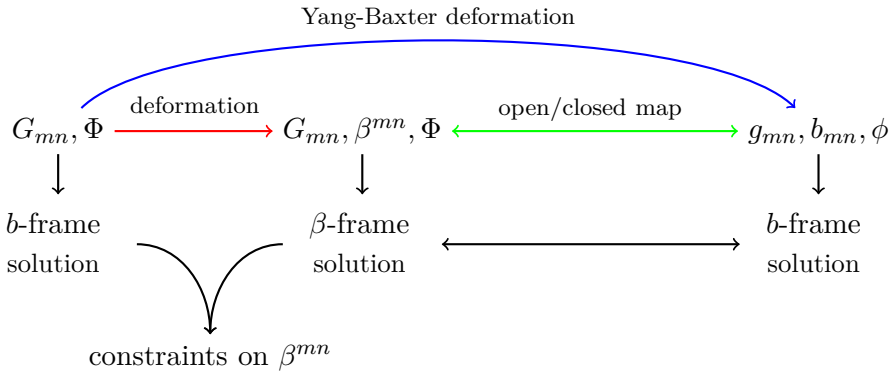
\begin{figure}[ht]
\centering
\begin{tikzpicture}
\node at (0,6) (Gf) {$G_{mn},\Phi$};
\node at (4,6) (GQf) {$G_{mn}, \beta^{mn}, \Phi$};
\node at (10,6) (gBf) {$g_{mn}, b_{mn}, \phi$};
\node at (0,4.5) (EOMs0) {\parbox{1.8cm}{\begin{tabular}{c}$b$-frame\\ \small solution\end{tabular}}};
\node at (4,4.5) (bEOMs) {\parbox{1.82cm}{\begin{tabular}{c}$\beta$-frame\\ \small solution\end{tabular}}};
\node at (10,4.5) (EOMs) {\parbox{1.8cm}{\begin{tabular}{c}$b$-frame\\ \small solution\end{tabular}}};
\node at (2,3) (cstr) {constraints on $\b^{mn}$};

\node at (1.8,6) [above=0.1cm] {\footnotesize deformation};
\node at (7,6) [above=0cm] {\footnotesize open/closed map};
\node at (5,7.5)  {\footnotesize Yang-Baxter deformation};

\draw[->, thick, red] (Gf) -- (GQf);
\draw[<->, thick, green] (GQf) -- (gBf);
\draw[->, thick] (Gf) -- (EOMs0);
\draw[->, thick] (GQf) -- (bEOMs);
\draw[->, thick] (gBf) -- (EOMs);
\draw[<->, thick] (EOMs) -- (bEOMs);

\path[->, thick] (EOMs0) edge[out=0, in=90] (cstr);
\path[->, thick] (bEOMs) edge[out=180, in=90] (cstr);
\draw[->, thick, blue] (Gf) .. controls (1.5,7.5) and (8.5,7.5) .. (gBf);

\end{tikzpicture}
\caption{
	Relationships between the relevant theories and their solutions. $b$-frame refers to the standard supergravity (possibly generalised), while $\beta$-frame is the theory of~\cite{Andriot:2013xca}. Yang-Baxter deformation acts within usual supergravity, but we interpret it as a composition of the open/closed string map with a deformation by $\beta^{mn}$. This leads to the constraints for $\beta^{mn}$ (essentially the CYBE) arising from supergravity field equations.
}
\end{figure}	

\subsection{\texorpdfstring{$\b$}{beta}-supergravity}

Dynamical fields of Double Field Theory (DFT) are the generalised metric $\mH_{MN}$ and the invariant dilaton $d$, both of which depend on the full doubled space-time with coordinates $\XX^M=(x^m,\tx_m)$. The generalised metric is an element of the coset space $O(10,10)/O(1,9)\times O(1,9)$ and can be parametrised by
\begin{equation}
\label{Hmetric0}
 \mH_{MN}=
 \begin{bmatrix}
   g_{mn}-b_{m}{}^l b_{ln} & & b_m{}^q \\
&\\
b_n{}^p & & g^{pq}
 \end{bmatrix}.
\end{equation}
The invariant dilaton is given by $d = \phi + \frac14 \log g$, with $g=\det g_{mn}$. The DFT action has been constructed in \cite{Hohm:2010pp} and can be written as
\begin{equation}
\label{Odd_action}
\begin{aligned}
    S_{HHZ}=\int dx\, d\tilde x\, e^{-2d}&\left(\fr18\mc{H}^{MN}\dt_M\mc{H}^{KL}\dt_{N}\mc{H}_{KL} -\fr12
\mc{H}^{KL}\dt_L\mc{H}^{MN}\dt_N\mc{H}_{KM} - \right.\\
      & \left.- 2\dt_M d\, \dt_N\mc{H}^{MN}+4\mc{H}^{MN}\dt_Md\, \dt_Nd \lefteqn{\ph{\fr12}}\right).
      \end{aligned}
\end{equation}
Integration over the full doubled space here is formal and just denotes that one has to integrate over all coordinates on which non-trivial dependence was left after solving the section constraint:
\begin{equation}
\h^{MN}\dt_M \otimes \dt_N = 0.
\end{equation}
This constraint is necessary for the algebra of generalised diffeomorphisms to close, and it means that all fields of the theory must depend only on one half of the total number of the coordinates. More precisely, one cannot have dependence on both a geometric coordinate $x$ and its dual $\tx$.

Solving the section constraint in the DFT action explicitly, one recovers the action for the bosonic sector of type II supergravity
\begin{equation}
S_{HHZ}\Big|_{SC}=\int dx\, e^{-2\ff}\sqrt{-g}\,\Big(\mR(g)-\fr{1}{12}H^{mnk}H_{mnk}+g^{mn}\dt_m\ff\dt_n\ff \Big),
\end{equation}
where $H=db$, and $\mR$ is the curvature scalar. This establishes the fact that the $b$-frame parametrisation of double field theory~\eqref{Hmetric0} is equivalent to the standard supergravity.

However, one is free to choose a different parametrisation for the generalised metric
\begin{equation}
\label{Hmetric1}
 \mH_{MN}=
 \begin{bmatrix}
   G_{mn} & & - \b_m{}^q\\
&\\
-\b_n{}^p & & G^{pq} - \b^p{}_l \b^{lq}
 \end{bmatrix}.
\end{equation}
These two natural choices of parametrization follow from the upper- and lower-triangular form of the generalised vielbein 
\begin{equation}
E^A_M=(E_0)^A_N (O_1)^N_M,\quad \text{or}\quad E^A_M=(E_0)^A_N(O_2)^N_M,
\end{equation}
where the matrices $E_{0}$ and $O_{1,2}$ are given by 
\begin{equation}
\begin{aligned}
&E_0=
\begin{bmatrix}
e^a_m & 0 \\
0 & e^n_b
\end{bmatrix}, &&
O_1=
\begin{bmatrix}
\d^m_k & -\b^{ml} \\
0 & \d^l_n
\end{bmatrix},&&
O_2=
\begin{bmatrix}
\d^k_p           & 0 \\
-B_{lp} & \d^q_l
\end{bmatrix}.
\end{aligned}
\end{equation}
One can obtain~\eqref{Hmetric0} and~\eqref{Hmetric1} as
$\mH_{MN}=E_M^{A} E_N^{B} \mH_{AB}$ with
\begin{equation}
\mH_{AB}=\begin{bmatrix}
\h_{ab} & & 0  \\ \\
0 & & \h^{ab}
\end{bmatrix}.
\end{equation}
The open/closed string map~\eqref{op-cl-map} in this context is a condition that~\eqref{Hmetric0} and~\eqref{Hmetric1} describe the same object.

Again solving the section constraint explicitly and substituting the generalised metric parametrised by $G_{mn}$ and $\b^{mn}$ one obtains the action of $\b$-supergravity as described in \cite{Andriot:2013xca}
\begin{equation}
\label{L_beta}
\tilde{L}_{\b} = e^{-2\Phi} \sqrt{-G}\ \bigg( \mR(G) + \cR +4(\dt \Phi)^2 -\frac{1}{2} R^2 + 4 (\b^{mp}\dt_p \Phi + I^m)^2 \bigg),
\end{equation}
where one defines
\begin{equation}
 \label{fluxes}
\begin{aligned}
 \cR&=G_{mn} \cR^{mn}, \quad \cR^{mn} = -\b^{pq}\dt_q \cG_p^{mn} + \b^{mq}\dt_q \cG_p{}^{pn} + \cG_p{}^{mn} \cG_q{}^{qp} - \cG_p{}^{qm} \cG_q{}^{pn},\\
 \cG_p^{mn}& = \frac{1}{2}G_{pq}\left(-\b^{mr}\dt_r G^{nq} - \b^{nr}\dt_r G^{mq} +  \b^{qr}\dt_r G^{mn} \right) + G_{pq} G^{r(m} \dt_r \b^{n)q}-\frac{1}{2} \dt_p \b^{mn} \\
 &=\na^{(m}\b^{n)}{}_p-\fr12 \na_p \b^{mn}+\b^{mq}\G^n{}_{pq} ,\\
 I^m &= \na_k \b^{km} \equiv -\cG_k{}^{km} = -\dt_k \b^{mk} + \frac{1}{2} \b^{mn} G_{pq} \dt_n G^{pq}  \  \\
 \cN^m V^p &= -\b^{mn} \dt_n V^p - \cG_n{}^{mp} V^n, \quad \cN^m V_p = -\b^{mn} \dt_n V_p + \cG_p{}^{mn} V_n \\\
R^{mnp}&= 3 \b^{q[m}\na_q \b^{np]} 
\end{aligned}
\end{equation}
For the Riemann and Ricci curvature tensors the standard conventions have been taken
\begin{equation}
\begin{aligned}
[\na_m,\na_n]V^p&=\mR^{p}{}_{q,mn}V^q,\\
\mR^{p}{}_{q,mn}&=2\dt_{[m}\G^p{}_{n]q}+2\G^p{}_{k[m}\G^k{}_{n]q},\\
\mR_{mn}&=\mR^p{}_{m,pn}.
\end{aligned}
\end{equation}

For further use $\cR^{mn}$ can be written explicitly in terms of the 2-vector $\b^{mn}$ as
\begin{equation}
\begin{aligned}
\cR^{mn}=&-\b^{pq}\na_q\na^{(m}\b^{n)}{}_p+\b^{mq}\na_q\na_p\b^{np}-\fr12 \b^{pq}\b^{rm}\mR^{n}{}_{r,pq}+\na^{(m}\b^{n)}{}_p\na_q\b^{pq}\\
&-\fr12 \na_p\b^{mn}\na_q\b^{pq}-\fr12 \na_q\b^{mp}\na_p \b^{nq}+\fr12\na^q\b^m{}_p\na_q \b^{pn}+\fr14 \na^m \b^{pq} \na^n \b_{pq}\\
&+\na^q\b^{p(m}\na^{n)}\b_{pq}.
\end{aligned}
\end{equation}
The tensor $R^{mnk}$ is the so-called non-geometric R-flux (when integrated over a non-trivial 3-cycle) which signals about non-associativity properties of a chosen background.  The vector $I^m$ is also a signature of non-geometry and at the level of Yang-Baxter deformations can be realised in generalised supergravity, as will be discussed below.

Equations of motion for the fields $G_{mn},\b^{mn}$ and $\Phi$ can be derived from the equations of motion of DFT upon the condition $\tdt^m\bullet =0 $ or by direct variation of the Lagrangian \eqref{L_beta}. These can be written as 
\begin{equation}
\label{full_EOMs}
\begin{aligned}
  \frac{1}{4} \left(\mR(G) + \cR(G) -\frac{1}{2} R^2 \right)= &\ (\dt \Phi)^2 - \nabla^2 \Phi + (\b^{mr}\dt_r \Phi + I^m)^2  \\
  & + G_{mn}\cN^m(\b^{nr}\dt_r \Phi + I^n);\\
  \mR_{pq} - \cR_{(pq)} + \frac{1}{4}  R_{pmn} R_q{}^{mn}= &  - 2 \nabla_p \dt_q \Phi  -2 \cN_{(p} \big( \b_{q)r} \na^r \Phi \big) -2 \cN_{(p} I_{q)} ; \\
 \frac{1}{2}   \left( e^{2\Phi} \cN^m (e^{-2\Phi}  R_{mrp}) + 2 I^m R_{mrp}\right)= & -\frac{1}{2}  e^{2\Phi} \na^m (e^{-2\Phi} \na_m \b_{rp} ) - 2  \mR_{[p}{}^{s} \b_{r] s}\\
 &+ e^{-2\Phi} \na^q ( e^{2\Phi}  \na_{[p} \b_{r]q}) + 4 G_{n[p} \na_{r]} (\b^{nq} \dt_q \Phi )  \ 
\end{aligned}
\end{equation}
Following the logic of \cite{Andriot:2013xca} these equations can be understood as equations of motion of the conventional supergravity rewritten in terms of the new fields. Hence, any solution of supergravity field equations after the open/closed string map must satisfy the above equations. Conversely, for a conventional supergravity solution with added field $\beta^{mn}$ and no $B$-field to also be a solution after the open/closed string map, it should satisfy the equations of motion of beta-supergravity.

\subsection{Proof}

We are now finally in a position to prove that the homogeneous CYBE is sufficient for the supergravity equations of motion to be satisfied by the deformed background. One can explicitly check the field equations of $\b$-supergravity~\eqref{full_EOMs}, using the bi-Killing ansatz for the $\b$-deformation
\begin{equation}
\label{bikil}
\b^{mn} = r^{ab} k_a^m k_b^n,
\end{equation}
as well as the fact that the undeformed metric and the dilaton satisfy the conventional field equations:
\begin{equation}
\label{conv}
\begin{aligned}
 \mR(G)  &= 4(\dt \Phi)^2 - 4\nabla^2 \Phi \\
 \mR_{pq}(G)&= - 2 \nabla_p \dt_q \Phi .
\end{aligned}
\end{equation}
In addition we assume $\b^{mn}\dt_n \Phi=0$ since the isometries of the original solution preserve the dilaton as well. Another constraint comes from the fact that although we are dealing with the beta-frame formulation, this is still conventional supergravity rather than the generalised. This implies that the vector $I^m=\nabla_k\b^{km}$ must be zero. Upon the bi-Killing ansatz this condition gives 
\begin{equation}
I^{m}=\na_k\b^{km}= k_a{}^n\partial_n k_b{}^m r^{ab}=\fr12 f_{ab}{}^c k_c{}^mr^{ab}=0.
\end{equation}
Finally, for the $R$-flux we have, using the ansatz:
\begin{equation}
	R^{mnk} = 3\, \b^{q[m}\na_q \b^{nk]} = 3\, k_a^m k_b^n k_c^k \,f_{de}{}^{[a} r^{b|d|} r^{c]e}.
\end{equation}
Thus the assumption that the CYBE holds implies the vanishing of the $R$-flux.
Taking all this into account, the  equations of motion of $\b$-supergravity boil down to
\begin{equation}
\label{eqns_beta}
\begin{aligned}
   0&= \cR^{(mn)} , \\
0  &  = \frac{1}{2} e^{2\Phi} \na^m (e^{-2\Phi} \na_m \b_{pr} ) + 2 G_{n[p} \mR_{r]m} \b^{nm} - e^{-2\Phi} \na^m ( e^{2\Phi}  \na_{[r} \b_{p]m})
\end{aligned}
\end{equation}
Equation for the dilaton trivially follows from the trace part of the first equation here. 

In the proof the following identities for the Killing vectors, usually referred to as the Kostant formula, will be used
\begin{equation}
\label{Kost}
\begin{aligned}
\nabla_m\nabla_p k^q &= \mR^q{}_{p,mn}k^n, \\
\Box k^m&= -\mR^m{}_nk^n.
\end{aligned}
\end{equation}

Let us start first with the equation for $\b^{mn}$, which is the second line above, and expand the covariant derivatives of products. This gives
\begin{equation}
\begin{aligned}
 0 & =  \frac{1}{2} e^{2\Phi} \na^m (e^{-2\Phi} \na_m \b_{pr} ) + 2  \mR_{[r}{}^{m} \b_{p]m}- e^{-2\Phi} \na^m ( e^{2\Phi}  \na_{[r} \b_{p]m}) \\
 & = - \na^m \Phi \na_m \b_{pr}+\frac{1}{2} \Box \b_{pr} + 2  \mR_{[r}{}^{m} \b_{p]m} - \na^m \na_{[r} \b_{p]m}-2\b_{m[p}\na_{r]}\na^m \Phi\\
  & = - \na^m \Phi \na_m \b_{pr}+\frac{1}{2} \Box \b_{pr} +  \mR_{[r}{}^{m} \b_{p]m} - \na^m \na_{[r} \b_{p]m}
\end{aligned}
\end{equation}
where we have used symmetry of the dilaton $\b^{mn}\na_n \Phi=0$ in the second line and the equation of motion for the dilaton  $\mR_{pq}(G)= - 2 \nabla_p \dt_q \Phi$ in the third line. Furthermore, the first term above can be transformed as follows
\begin{equation}
\begin{aligned}
- \na_m \Phi \na^m \b^{pr}& = -2 r^{ab} \na_m \Phi \, \big(\na^m k_b{}^{[r}\big)k_a{}^{p]} =
2 r^{ab} \na_m \Phi \, k_a{}^{[p}\na^{r]} k_b{}^{m}\\
&= 2 r^{ab} k_a{}^{[p}\na^{r]}\big( k_b{}^{m} \na_m \Phi\big) - 2r^{ab} k_b{}^{m} k_a{}^{[p} \na^{r]}\na_m \Phi = \mR^{[r}{}_m\b^{p]m}.
\end{aligned}
\end{equation}
Hence, the equation of motion simplifies and can be rewritten more conveniently as
\begin{equation}
\label{eqn_beta_fin}
0= \fr12\Box \b^{pq} + \na_m \na^{[p} \b^{q]m}-2\mR^{[p}{}_m\b^{q]m}
\end{equation}
Let us show, that this vanishes upon $I^m=0$. We start with the second term above, which can be manipulated as follows
\begin{equation}
\begin{aligned}
\na_m \na^{[p} \b^{q]m}&=[\na_m, \na^{[p}] \b^{q]m}=\mR^{[q}{}_{r,m}{}^{p]}\b^{rm}+\mR_r{}^{[p}\b^{q]r}.
\end{aligned}
\end{equation}
For the first term in \eqref{eqn_beta_fin} we write (antisymmetry in $\{p,q\}$ is always understood)
\begin{equation}
\begin{aligned}
\fr12 \Box \b^{pq}&=\na^k\big(k_a{}^{p}\na_k k_b{}^{q}\big)r^{ab}=
\na^k k_a{}^{p}\na_k k_b{}^{q}r^{ab}+k_a{}^{p}\Box k_b{}^{q}r^{ab}\\
&=-\na^p \big(k_a{}^{k}\na_k k_b{}^{q}\big)r^{ab}+ k_a{}^{k}\na^p\na_k k_b{}^{q}r^{ab}+k_a{}^{p}\Box k_b{}^{q}r^{ab}\\
&=-\fr12\na^{[p} k_c{}^{q]}f_{ab}{}^cr^{ab}+ k_a{}^{k}\mR^{[q}{}_{k,}{}^{p]}{}_rk_b{}^{r}r^{ab}-k_a{}^{[p}\mR^{q]}_rk_b{}^{r}r^{ab}\\
&=-\mR^{[q}{}_{k,r}{}^{p]}\b^{kr}+\mR^{[p}{}_r\b^{q]r},
\end{aligned}
\end{equation}
where we have used the Killing equation $\na^{(m}k_a{}^{n)}=0$  in the second line,  the Kostant formula \eqref{Kost} in the third line, and symmetry properties of the Riemann tensor in the last line. Substituting everything back to \eqref{eqn_beta_fin} one concludes that the equations of motion for $\beta^{mn}$ are satisfied identically on the bi-Killing ansatz, given $I^m=0$.

Consider now the Einstein-Hilbert equation, which is the first line in \eqref{eqns_beta}. For this we need to write the tensor $\cR^{mn}$ and the corresponding connection symbols in terms of the field $\b^{mn}$. Let us start with splitting the connection symbols into tensorial and non-tensorial part
\begin{equation}
\begin{aligned}
\cG_p{}^{mn}&=\bar{\G}_p{}^{mn}+\g_p{}^{mn},\\
\bar{\G}_p{}^{mn}&=-k_a{}^m\nabla_p k_b{}^n r^{ab},\\
\g_p{}^{mn}&=\b^{mq}\G^n{}_{pq}
\end{aligned}
\end{equation}
Substituting this decomposition back into $\cR^{mn}$ and rewriting everything in terms of covariant derivatives one obtains the following most general expression
\begin{equation}
\begin{aligned}
\cR^{mn}&=-\b^{pq}\nabla_q \bar{\G}_p{}^{mn}+\b^{mq}\nabla_q \bar{\G}_p{}^{pn}+\bar{\G}_p{}^{mn}\bar{\G}_q{}^{qp}
-\bar{\G}_p{}^{qm}\bar{\G}_q{}^{pn}\\
&+\fr12 \b^{pq}\b^{mr}\mR^n{}_{r,pq}
\end{aligned}
\end{equation}
Substituting the bi-Kiling ansatz here we get
\begin{equation}
\label{R_k}
\cR^{mn}=k_c^q\nabla_q k_b{}^m k_a{}^n f_{ef}{}^a r^{be}r^{fc}-\fr12 \b^{mq}\nabla_q k_c{}^n f_{ab}{}^c r^{ab}+\fr12 k_a{}^m\nabla_p k_b{}^n k_e{}^p f_{cd}{}^e r^{ab}r^{cd}.
\end{equation}
The last two terms here have similar structure and vanish upon $f_{ab}{}^cr^{ab}=0$, while the first can be shown to be proportional to the CYBE plus terms of the same structure as the last two. Indeed, consider the following expression proportional to the CYBE
\begin{equation}
\begin{aligned}
-3 k_c{}^q \nabla_q k_b{}^m k_a{}^n \big(f_{ef}{}^{[a}r^{b|e|}r^{c]f}\big)
=&-k_c{}^q \nabla_q k_b{}^m k_a{}^n f_{ef}{}^{a}r^{[b|e|}r^{c]f}-2k_c{}^q \nabla_q k_b{}^m k_a{}^n f_{ef}{}^{[c}r^{b]f}r^{ae}\\
=&-k_c{}^q \nabla_q k_b{}^m k_a{}^n f_{ef}{}^{a}r^{[b|e|}r^{c]f}-k_g{}^m k_a{}^nf_{cb}{}^g  f_{ef}{}^{c}r^{bf}r^{ae}\\
=&-k_c{}^q \nabla_q k_b{}^m k_a{}^n f_{ef}{}^{a}r^{be}r^{cf}+\fr12k_g{}^m k_a{}^nf_{ce}{}^g  f_{fb}{}^{c}r^{bf}r^{ae}
\end{aligned}
\end{equation}
where in the second line we used the bracket of two Killing vectors and in the third line we used the Jacobi identity for the structure constants $f_{ab}{}^c$. The first term in the last line above is precisely the one we have found in $\cR^{mn}$, while the second vanishes upon $f_{ab}{}^c r^{ab}=0$, and the equation $\cR^{mn}=0$ is proportional to the CYBE given the condition $I^m=0$.

The above calculations provide a proof of the conjecture of~\cite{Bakhmatov:2017joy}: we have shown that the classical Yang-Baxter equation emerges from the open/closed string map, applied to supergravity solutions deformed by a 2-form~\eqref{bikil}, assuming that we do not have to deal with generalised supergravity $\na_m\b^{mn}=0$.

\section{Discussion}
\label{sec3}

In this work we have provided a detailed proof that the deformation of a supergravity solution parametrised by tensor $\b^{mn}$ subject to the bi-Killing ansatz
\begin{equation}
\b^{mn} = r^{ab} k_a{}^m k_b{}^n
\end{equation}
is a solution of supergravity equations, if the $r$-matrix $r^{ab}$ satisfies classical Yang-Baxter equation and the vector $I^m=\na_n\b^{mn}=0$ vanishes. The idea of the proof is that the field equations of supergravity for the deformed background $g_{mn}, b_{mn}, \phi$ obtained by the open-closed string map 
\begin{equation}
(G^{-1} + \b)^{-1} = g + b
\end{equation}
are equivalent to the equations of motion of $\b$-supergravity for the background $G_{mn}$, $\b^{mn}$, $\Phi$. The fact that the background before the deformation $G_{mn}, \Phi$ satisfies the equations of motion of the conventional supergravity provides dynamical equations for the deformation parameter $\b^{mn}$. 

The described approach provides a natural framework for addressing deformations of supergravity backgrounds both integrable and not. There are several directions where this approach can be extended and which we consider interesting.

\subsection{Applications and outlook}

\paragraph{General deformations: the case of flat backgrounds.} Starting with a given solution of supergravity equations of motion $G_{mn}, \Phi$, consider its general bi-vector deformation
\begin{equation}
G^{mn} \to G^{mn}+\b^{mn}.
\end{equation}
According to the above logic, equations of motion of beta-supergravity for this background provide constraints which must be satisfied for the deformation $\b^{mn}$ to generate a solution. Hence, beta-supergravity allows to classify all such deformations for a given background, not only of the Yang-Baxter form, however, not distinguishing between integrable and non-integrable deformations.

As a simple illustration of the idea consider the flat background of the form
\begin{equation}
\begin{aligned}
G_{mn}=\h_{mn}, && \Phi=0.
\end{aligned}
\end{equation}
Equations of motion of beta-supergravity then take the following form
\begin{equation}
\label{flat_EOMs}
\begin{aligned}
 \cR -\frac{1}{2} R^2 =&\ 4( I^m)^2+ 4\cN^m I_m \\
   \cR_{(pq)} - \frac{1}{4}  R_{psv} R_{q}{}^{sv} =&\   2  \cN_{(p} I_{q)} \\
 \cN^m   R_{mrp} + 2 I^m R_{mrp}    = &\ \dt^m  \dt_m \b_{pr} + 2 \dt_{[r} I_{p]}   .
\end{aligned}
\end{equation}
Any solution of these equations for $\b^{mn}$ will give a deformation which generates a background $g_{mn}, b_{mn}, \phi$ that solves field equations of supergravity. Note that at this stage we do not restrict ourselves to the case $I^m=0$.

To provide explicit examples, let us consider the most simple case of dimension $d=2$. This may be understood as a class of backgrounds where deformed is only a two-dimensional block inside the full metric. For this case the deformation parameter is a single function $\b^{mn}=\b \e^{mn}$, where the alternating symbol is defined as $\e^{mn}\e_{nk}=\a\d^m{}_k$. Here $\a=+1$ for  the Minkowski case and  $\a=-1$ for the Euclidean case. In addition $R^{mnk}=0$ as there is no R-flux in two dimensions.

With these  simplifications the equations for the deformation parameter become
\begin{equation}
\begin{aligned}
\cR^{mn}&=- \beta \alpha {\partial}^{m n}{\beta}\,  + \beta {\epsilon}^{m k} {\epsilon}^{n l} {\partial}_{k l}{\beta}\,  + \frac{1}{2}\, \alpha {\partial}^{m}{\beta}\,  {\partial}^{n}{\beta}\,  - \frac{1}{2}\, {\epsilon}^{m k}{\epsilon}^{n l} {\partial}_{k}{\beta}\,   {\partial}_{l}{\beta}\,  + \frac{1}{2}\, \alpha {\delta}^{m n} {\partial}^{l}{\beta}\,  {\partial}_{l}{\beta}\, \\
I^m&=\e^{km}\dt_k \b.
\end{aligned}
\end{equation}
 This implies 
\begin{equation}
\begin{aligned}
4I^2- 2\, {\partial}^{m}{{\beta}^{n k}}\,  {\partial}_{n}{{\beta}_{m k}}\,  -\, {\partial}^{m}{{\beta}^{n k}}\,  {\partial}_{m}{{\beta}_{n k}} =0,
\end{aligned}
\end{equation}
which is identically satisfied in the dimension chosen. To get to explicit solutions of these equations one has to consider the Minkowski and Euclidean cases separately. Let us start with the former, where $\a=+1$ and we define $\e^{01}=1$. This gives the following equations
\begin{equation}
\begin{aligned}
\b\ddot{\b}+\dot{\b}^2&=0,\\
\b\dot{\b}'+\dot{\b}\b'&=0,\\
\b\b''+\b'{}^2&=0
\end{aligned}
\end{equation}
The first equation implies $\b(t,x)=\pm\sqrt{a(x)t+b(x)}$, from the second equation one concludes $a'(x)=0$ and from the third one that $b''(x)=0$. Altogether this gives the following general solution of the above equations
\begin{equation}
\b(t,x)=\pm\sqrt{a t+b x+c},
\end{equation}
where $a,b,c$ are constant parameters.

For the Euclidean case one has $\a=-1$ and the equations become
\begin{equation}
\begin{aligned}
\b\dt_{11}{\b}+(\dt_1{\b})^2&=0,\\
\b\dt_{12}{\b}+\dt_1{\b}\dt_2\b&=0,\\
\b\dt_{22}\b+(\dt_2\b){}^2&=0.
\end{aligned}
\end{equation}
Following the same steps as above one obtains a similar solution
\begin{equation}
\b(x^1,x^2)=\pm\sqrt{a x^1+b x^2+c}.
\end{equation}
Altogether the solutions for the Minkowski and Euclidean cases can be written as
\begin{equation}
\b=\pm\sqrt{k_m x^m+c},
\end{equation}
where $k^m$ is a constant vector in two dimensions.

Performing the open/closed string map, we obtain the following deformation of the initial flat background
\begin{equation}
g=\fr{1}{c + k_p x^p}\h, \quad b=\pm\fr{\sqrt{k_m x^m+c-1}}{c+k_m x^m}
\begin{bmatrix}
0 & -1 \\
1 & 0
\end{bmatrix}, \quad
 e^{-2(\phi-\phi_0)}=|c+k_mx^m|.
\end{equation}
This $b$-field is trivial and can be written as a pure gauge $b_{mn}=\dt_{[m}\a_{n]}$ with
\begin{equation}
\a_m=\mp 4\fr{\e_{mn}k^n}{k^2}\Big[\sqrt{c-1+k_p x^p}+\arctan\big(\sqrt{c-1+k_p x^p}\big)\Big].
\end{equation}
The background itself is conformally flat and in general is non-trivial.

It is important to note, that the obtained deformation is not of the Yang-Baxter class and cannot be represented in the form of the bi-Killing ansatz. In order to recover a bi-Killing $\beta^{mn}$, one would need to solve the $\beta$-supergravity version of generalised supergravity equations, since $I^m=\na_n\b^{nm}\neq 0$. This example illustrates, that the described approach is able to produce deformations of a very general class.

\paragraph{Deformations with non-vanishing $I^m$.} The proof here has been restricted to the case $I^m=\na_k\b^{km}=0$. We know, however, that CYBE similarly arises from generalised supergravity equations of motion after the open/closed string map, if $I^m$ is non-zero. In principle one could seek to find a generalisation of $\b$-supergravity, such that the open/closed string map would take this new theory to generalised supergravity. 

The generalised supergravity field equations can be obtained from the usual ones by replacing
\begin{equation}
	\dt_m \phi \rightarrow X_m  = \dt_m \phi + (g-b)_{mn} I^n =  \dt_m \phi + (g-b)_{mn} \nabla_k \b^{kn}.
\end{equation}
However, the natural conjecture that the same replacement can be done in the $\beta$-super\-gra\-vity field equations does not result in correct equations of motion, e.g. one cannot reproduce the bi-Killing ansatz for $\b^{mn}$ by solving them. Hence, a more subtle deformation of equations of motion of beta-supergravity needs to be found, which is still an open question. An earlier generalisation of the DFT to include generalised supergravity has appeared in~\cite{Sakatani:2016fvh, Baguet:2016prz}.

\paragraph{Deformations of backgrounds with non-zero $b$-field.} As was already discussed, $\b$-supergravity descends from the DFT after a special parametrisation of degrees of freedom of the generalised metric is chosen. This choice of parametrisation allows to consider deformations of backgrounds with non-vanishing $b$-field using the same approach as above.

Consider generalised vielbein $E^M_A \in O(d,d)$ that includes both the 2-form Kalb-Ramond field 
$b_{mn}$ and the bivector $\b^{mn}$. The generalised vielbein  can be represented as a product of the following 
\begin{equation}
\begin{aligned}
 E^A_M&=(E_0)^A_N (O_1)^N_K(O_2)^K_M, \\
 E'^A_M&=(E_0)^A_N(O_2)^N_K(O_1)^K_M,\\
\end{aligned}
\end{equation}
where the matrices $E_{0}$ and $O_{1,2}$ are given by 
\begin{equation}
\begin{aligned}
&E_0=
\begin{bmatrix}
e^a_m & 0 \\
0 & e^n_b
\end{bmatrix}, &&
O_1=
\begin{bmatrix}
\d^m_k & -\b^{ml} \\
0 & \d^l_n
\end{bmatrix},&&
O_2=
\begin{bmatrix}
\d^k_p           & 0 \\
-B_{lp} & \d^q_l
\end{bmatrix}.
\end{aligned}
\end{equation}
Naively, the corresponding generalised metric contains more degrees of freedom than the conventional 
generalised metric of DFT. However, equating generalised metric in each of these parametrizations to generalised metric in the conventional parametrization, one obtains the following simple field redefiniton rules respectively
\begin{equation}
\label{redef}
\begin{aligned}
 (G^{-1}+\b)^{-1}+B&=g+b,\\
 (G-B)^{-1}-\b&=(g-b)^{-1}.
\end{aligned}
\end{equation}
The Double Field Theory equations of motion for these parametrizations provide a natural framework for generating deformations of backgrounds with non-vanishing $B$.

\paragraph{Deformations of $d=11$ backgrounds within exceptional field theory.} For backgrounds of $d=10$ supergravity one introduces the deformation parameter $\b$ and then performs the open/closed string map 
\begin{equation}
(G^{-1}+\b)^{-1}= g+b.
\end{equation}
This map, when understood as a change of frame in supergravity, imposes the CYBE as a consistency condition on the parameter $\b$, which ensures that the deformed background satisfies the field equations given that the background $G_{mn}, \Phi$ is a solution.

From the point of view of the double field theory these are just two different ways to write the generalised metric $\mH_{MN}$, which is an element of the coset $O(d,d)/O(d)\tm O(d)$.

The same idea can be applied to solutions of $d=11$ supergravity. In this case one should employ the generalised metric of exceptional field theory~\cite{Berman:2011jh,Berman:2011cg}, which transforms under U-duality and contains fields $g_{\m\n}, C_{\m\n\r}, C_{\m\n\r\s\k}, \dots$. Acting in a similar fashion one can change the frame by a U-duality rotation, so as to have $G_{\m\n}, \W^{\m\n\r}, \W^{\m\n\r\s\k}, \dots$ as the fundamental degrees of freedom. Imposing that the metric $G_{\m\n}$ be a solution to the conventional $d=11$ equations of motion, the parameters $\W$ are deformations which generalise the non-commutativity parameter $\b$. One could speculate that a tri-Killing ansatz will be appropriate,
\begin{equation}
\W^{mnk}=k_a{}^mk_b{}^nk_c{}^k \r^{abc},
\end{equation}
with some totally antisymmetric tensor $\r^{abc}$. We remark that recently a tri-vector supergravity deformations have been considered in the framework of generalised geometry~\cite{Ashmore:2018npi}. Substituting this deformation back into the equations of motion of exceptional field theory in the $\W$-frame, and assuming the $d=11$ supergravity equations of motion for the original background, will result in some algebraic equations for the $\r$-tensor. The significance and nature of these equations at present can be debated, but functionally they arise in the same manner, as the CYBE in the case of $d=10$ deformations. Detailed investigation of this and related issues is a promising direction for the future work.

\acknowledgments

We would like to thank David Andriot, Sadik Deger, Eoin \'O Colg{\'a}in, and Shahin Sheikh-Jabbari for valuable discussions. EtM would like to thank APCTP and personally Eoin \'O Colg{\'a}in for warm hospitality. The work of EtM is supported by the Russian state grant Goszadanie 3.9904.2017/8.9 and by the Foundation for the Advancement of Theoretical Physics and Mathematics ``BASIS''. The research of IB was facilitated by the Ministry
of Science, ICT \& Future Planning, Gyeongsangbuk-do and Pohang City. 
This research was partially supported by the program of competitive growth of Kazan Federal University.

\providecommand{\href}[2]{#2}\begingroup\raggedright\endgroup

\end{document}